\documentclass[a4paper, oneside, reqno, 12pt]{article}
 
\usepackage[a4paper, top=1.1in, bottom=20ex, left=30mm,
right=30mm,headheight=28pt,headsep=8pt]{geometry}
 \usepackage[super]{natbib}
\usepackage{graphicx}
\usepackage{epsfig}
\usepackage{amsthm}
\usepackage{amsmath}
\usepackage{latexsym}
\usepackage{amsfonts}
\usepackage{amssymb}
\usepackage{hyperref} 
\usepackage{float}       
\usepackage{color}

\begin{document}

 \title{\LARGE \textbf{Connecting the grain-­shearing mechanism of wave propagation in marine sediments to fractional order wave equations}}
 \author{{Vikash Pandey\footnote{Corresponding author. Electronic address: vikashp@ifi.uio.no} and
Sverre Holm}\\ 
\footnotesize   \emph{Department of Informatics, University of Oslo, P.O. Box 1080, NO-0316 Oslo, Norway}\\}

\date{December 20, 2016}

\maketitle

 \begin{abstract}

The characteristic time-dependent viscosity of the intergranular pore-fluid in Buckingham's grain-shearing (GS) model [Buckingham, J. Acoust. Soc. Am. \textbf{108}, 2796--2815 (2000)] is identified as the property of rheopecty. The property corresponds to a rare type of a non-Newtonian fluid in rheology which has largely remained unexplored. The material impulse response function from the GS model is found to be similar to the power-law memory kernel which is inherent in the framework of fractional calculus. The compressional wave equation and the shear wave equation derived from the GS model are shown to take the form of the Kelvin-Voigt fractional-derivative wave equation and the fractional diffusion-wave equation respectively. Therefore, an analogy is drawn between the dispersion relations obtained from the fractional framework and those from the GS model to establish the equivalence of the respective wave equations. Further, a physical interpretation of the characteristic fractional order present in the wave equations is inferred from the GS model. The overall goal is to show that fractional calculus is not just a mathematical framework which can be used to curve-fit the complex behavior of materials. Rather, it can also be derived from real physical processes as illustrated in this work by the example of grain-shearing.

 \medskip


\vspace{5pt}

\texttt{The peer-reviewed version of this paper is published in\\
\textit{The Journal of the Acoustical Society of America, Vol.~140, No 6, pp.~4225-4236, 2016.
\\ DOI: \href{http://dx.doi.org/10.1121/1.4971289}{http://dx.doi.org/10.1121/1.4971289}} 
\\ The complete manuscript is available online at \\
\textit{\href{http://asa.scitation.org/doi/full/10.1121/1.4971289}{http://asa.scitation.org/doi/full/10.1121/1.4971289}}
\vspace{3pt}
\\
\textcolor{blue}{\textbf{The current document is an e-print which differs in e.g.  pagination, reference
numbering, and typographic detail.}}}

 \end{abstract}

 \maketitle

 \vspace*{-16pt}


 \section{INTRODUCTION}

 \setlength{\parindent}{5ex}
 
Buckingham's grain-shearing (GS) model\citep{Buckingham2000}${}^,$\citep{Buckingham2005} of wave propagation in fluid-saturated, unconsolidated marine sediments is investigated in this article. The development of the model can be traced back to the earlier works of Buckingham.\citep{Buckingham1997}${}^,$\citep{Buckingham1998} In comparison to the classical Biot theory, the GS model assumes the absence of an elastic frame and shows the generation of both compressional and shear waves from intergranular sliding. The central idea of the model is that random stick-slip motion drives the intergranular shearing mechanism. Further, the shearing is mediated by a pore-fluid present in between the grains. The pore-fluid is characterized by a time-dependent viscosity $\xi\left(t\right)$ which is expressed as (see Eq.~(17) in Ref.~\citenum{Buckingham2000}):
\begin{equation}\label{eq01}
\xi\left(t\right)\approx\xi_{0}+\theta t,
\end{equation}
where $t$ is the time, and the zero-order term $\xi_{0}$ is the viscosity of the pore-fluid before the sliding is triggered. Since, the apparent viscosity of the pore-fluid increases with the duration of shearing, it makes the inter-granular shearing difficult to sustain. Therefore, Buckingham refers to this mechanism as ``strain hardening'', and the first-order term $\theta$ as strain-hardening coefficient. Although the term ``strain hardening'' or ``work hardening'' has sometimes been used in a similar context,\citep{Lomnitz1956}${}^,$\citep{ZhangZhou2016} we prefer to reserve it to the field of material science where it is commonly used to explain the increase in hardness of materials in the form of yield strength when stretched beyond the linear elastic region, i.e., the non-Hookean behavior.\citep{Callister2012}

\medskip

The time-dependent increase of viscosity of the pore-fluid expressed by Eq.~(\ref{eq01}) is essentially a shear-thickening property and termed as rheopecty (or, rheopexy) in the field of non-Newtonian rheology.\citep{Mewis2009}${}^,$\citep{Deshpande2010} A simple example of the rheopectic property can be observed in whipped cream; the longer one whips, the thicker it gets. Other examples are suspensions containing a mixture of clay, calcium carbonate, starch, and water\citep{Yannas1961} and pharmaceutical products.\citep{Dewar2006} Compared to the rheopectic behavior its opposite time-dependent shear-thinning property called thixotropy is quite common. Some examples of thixotropic fluids are paint, honey, coal-water slurries, cement paste, and waxy crude oil. The apparent increase and decrease in viscosity is often due to the building-up and breaking down of internal structures which depend on the kinematic shearing history or ``memory'' of the fluid.\citep{Mewis2009}${}^,$\citep{Deshpande2010} The rheopectic and thixotropic properties must not be confused with the time-independent dilatant and pseudoplastic behavior which are yet another category of non-Newtonian fluids and depend on shear rate (or shear stress).\citep{Mewis2012}${}^,$\citep{Huang2015} In light of these arguments, we will refer to $\theta$ in Eq.~(\ref{eq01}) as the \textit{rheopectic coefficient} of the pore-fluid.  

\medskip

In contrast to Buckingham's use of a time-varying Maxwell model to describe the rheopectic behavior of GS mechanism in marine sediments,\citep{Buckingham2000} the time-dependent behavior of complex media has often been modeled using fractional viscoelastic models.\citep{Zhou2011}${}^-$\citep{Nadzharyan2016} The key element in the fractional models is a fractional dashpot, the constitutive stress-strain relation of which is given as,\citep{Mainardi2010}
\begin{equation}\label{eq02}
\chi=E_{0}\tau_{\chi}^{\gamma}\frac{d^{\gamma}\varepsilon}{dt^{\gamma}},  
\end{equation}
where $\chi$ is the stress, $\varepsilon$ is the strain, $E_{0}$ is the elastic modulus at zero frequency, $\tau_{\chi}$ is the characteristic retardation time constant, and $0<\gamma<1$ is the order of the fractional derivative.

\medskip

This paper which is an expanded version of\citep{Holm2016} builds on Buckingham's GS model, and connects the physical mechanism of grain-shearing with the mathematical framework of fractional calculus. As we will later show in the paper, the wave equations obtained from the GS model when analyzed in the fractional framework, appear as the Kelvin-Voigt fractional derivative wave equation and the diffusion-wave equation.

\medskip

The motivation behind this study is two-fold. First, since fractional derivatives are not constrained to integer-orders, they predict power-law attenuation of the wave proportional to $\omega^{\gamma}$, where $\omega$ is the angular frequency of the wave, and the exponent $\gamma$ can be any real positive number. The apparent good fit of the anomalous phase-velocity dispersion curve and power-law attenuation curve predicted from the GS model with the experimental data is also the main characteristic of materials modeled using fractional calculus.\citep{Nasholm2011} Although, the curve-fitting agreement is mostly in the range of $10$-$400$ kHz,\citep{Buckingham2007}${}^,$\citep{Buckingham2014} the similarity between the material impulse response function (MIRF) from the GS model and the relaxation modulus of the fractional dashpot provides enough impetus to study the GS model in light of the fractional calculus. Also, fractional derivatives have lately been applied to model frequency dependent power-law wave attenuation in sediments.\citep{Maestas2016} 

\medskip

Second, we have recently demonstrated that just as non-Hookean materials lead to nonlinear acoustics, time-dependent non-Newtonian materials tend to power law behavior.\citep{Pandey2016} Interestingly, power-law characteristics are common in materials modeled using fractional derivatives, such as biological media\citep{Kohandel2005}${}^-$\citep{Zhang2016} and earth materials.\citep{Carcione2002}${}^,$\citep{Zhu2014} The time-dependent non-Newtonian media have remained little understood despite their importance to industrial applications.\citep{Deshpande2010}${}^,$\citep{Blakey2003} The underlying reason is the complexities in the associated physical processes, and therefore most models put forward for their investigation are often empirical.\citep{Mewis2009}${}^,$\citep{Mujumdar2002} 

\medskip

Recently, the use of the fractional derivative was shown in the modeling of time-dependent behavior of muddy clay,\citep{Yin2012} however fractional derivative was introduced ad-hoc without providing any satisfying physical justification. It should be noted that an arbitrary substitution has its drawbacks as sometimes it could lead to physically meaningless results. This limitation can be traced back to the way fractional order wave equations are derived from ad-hoc phenomenological models comprising combinations of springs and dashpots.\citep{Mainardi2010} Consequently, the parameter $\gamma$ which also corresponds to the order of the resulting fractional differential equations is usually estimated by curve-fitting the experimental data with the theoretically predicted curves.\citep{Zhou2011}${}^-$\citep{Nadzharyan2016}${}^,$\citep{Kohandel2005}${}^-$\citep{Zhang2016} On the one hand, fractional derivatives impart greater flexibility to the fitting process, yet their applications have remained restricted due to a lack in their physical interpretation.\citep{Podlubny2002}${}^,$\citep{Machado2015} On the other hand, in our recent publication,\citep{Pandey2016} we have interpreted the order in terms of the parameters of the viscoelastic model. Through this work, we want to contribute to a change in motivation behind the application of fractional derivatives from inductive to deductive.

\medskip

We have the opinion that although the foundation of this bridging has been laid by Buckingham in the form of the derived MIRF of power-law form, he may not have realized this connection himself when he formulated the GS model. Apparently it also seems that the use of the term ``strain-hardening'' hid the implications from the GS model to both non-Newtonian rheology and fractional calculus. This is evident from the fact that neither the connection to past work on power-law behavior, nor to fractional derivatives was mentioned. One of the possible reasons could be that sediment acoustics and non-Newtonian rheology haven't shared a strong connection before which is also witnessed from the lack of cross-referencing between the two fields. Another possibility is that the fractional community is comparatively smaller in size, and has been active only in the last thirty years or so. Moreover, their focus has mostly been on pure mathematical research paying little attention to its physical applications.

\medskip

The article is organized as follows. In Sec.~II, we provide a short review of Buckingham's GS model mainly outlining its underlying physical mechanism. Then, in Sec.~III, the framework of fractional calculus is introduced, followed by the derivation of the Kelvin-Voigt fractional derivative wave equation and the time-fractional diffusion-wave equation. The compressional and shear wave equations obtained from the GS model in Sec.~II are then mapped into the domain of fractional calculus in Sec.~IV. Also, the dispersion relations obtained in the fractional framework are shown to be equivalent to their respective counterparts from the GS model. Finally, in Sec.~V, we discuss the implications of this work.

\medskip

\section{GRAIN-SHEARING MODEL}

Buckingham's\citep{Buckingham2000} GS model takes into account the non-Biot dissipative mechanisms of grain-shearing in saturated, unconsolidated marine sediments. Since at the microscopic level, the grains come into contact but lack bonding in the conventional sense, the dynamic frame bulk and shear moduli are zero.\citep{Jackson2007} As illustrated in Fig.~1, the static overburden pressure develops micro-asperities at the contact surfaces of the sediment grains. As a result of this micro-roughness, the participating grains exhibit stick-slip motion when triggered by a velocity gradient set up by the initial wave disturbance. The intergranular sliding is mediated through the saturating pore-fluid present between the grains. The pore-fluid is characterized by a time-varying viscosity given by Eq.~(\ref{eq01}). Consequently, as the grains slide against each other, the generated viscous drag force increases as time progresses, making the intergranular sliding motion difficult to sustain. This attribute of the drag force manifested as a result of the rheopectic behavior of the pore-fluid is represented by a time-varying viscous dashpot in the Maxwell element. Further, compressional and translational shearing occurring along the radials of the circle of contact of the sliding grains finally build up as compressional and shear waves respectively. 

\begin{figure}[H]
\centering
\hspace*{0cm}
\includegraphics[scale=0.35]{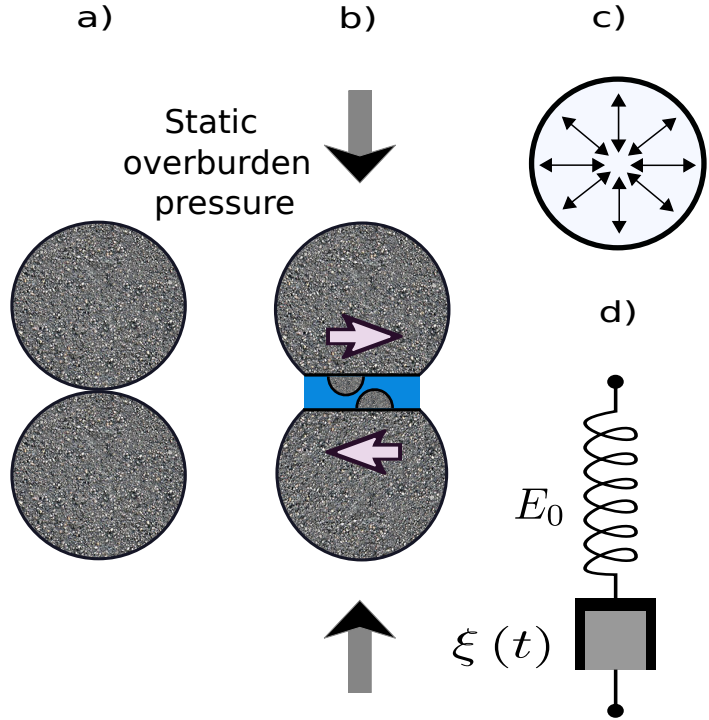}

\caption{(Color online)  Grain-to-grain shearing mechanism: a) Two spherical, saturated mineral grains in light contact. b) Deformed grains due to the static overburden pressure, development of micro-asperities (small solid hemispheres) separated by a thin film of pore-fluid, and intergranular tangential and compressional shearing. c) Radials of the circle of contact of the grains. d) Equivalent modified Maxwell model consisting of a series combination of a Hookean spring $E_{0}$ and a time-dependent viscous dashpot $\xi\left(t\right)$. Figure adapted from Ref.~\citenum{Buckingham2000}.}
\end{figure}

\medskip

Buckingham goes back to tribological properties to explain the rheopectic behavior of the pore-fluid where the properties emerge as a result of narrow confinement of the fluid in between the asperities. While this is a subject of ongoing research,\citep{Chotiros2014} we have the opinion that instead of anticipating ``regular'' seawater trapped in between the grains, a concentrated colloidal suspension is more likely in a marine-like environment. We would also like to add that rheopectic properties could occur due to changes in the state of aggregation or coagulation of the suspensions. Such behavior has already been observed in granular and clay-like colloidal suspensions in which the applied shear forces build up the interlinking of the structural elements giving rise to the apparent increase in viscosity.\citep{Yannas1961}${}^,$\citep{Mewis2012}${}^,$\citep{Wang1994} The build-up of the structures is often disrupted upon the removal of shear which makes the process reversible. The opposite happens in the case of thixotropic fluids. It should also be mentioned that the apparent change in viscosity cannot keep on increasing or decreasing forever. It must stop once a state of dynamic equilibrium is reached as observed in time-dependent fluid flow. Further, due to its complexity, the time-dependent non-Newtonian behavior is quite difficult to understand and so to model.

\medskip

It is also worthwhile to mention that contrary to the GS model in which rheopectic properties emerge from the biphasic granular-pore-fluid medium, time-dependency is already inherited in the stick-slip processes occurring in dried granular materials too. In the latter case it manifests itself as a result of the microscopic junctions formed due to the sliding of micro-asperities against each other.\citep{Dieterich1978}${}^,$\citep{Thogersen2014} Such situations might arise in partially saturated sediments as well as where the grains could come into direct contact with each other, and still exhibit time-dependent behavior.

\medskip

Further, given the fact that the material is granular and unconsolidated, for a given sample of randomly distributed grains of varying shapes, sizes and orientations, rotation of grains could also occur. This gives possibilities for squeezing of the pore-fluid in between the grains which could finally emerge as a squirt flow similar to the predictions from the Biot--Stoll plus grain-contact squirt and shear flow (BICSQS) model.\citep{Chotiros2004}

\medskip

Although in this article we have focused on the GS model only, it should also be mentioned that the model has undergone two improvements in the last fifteen years. The first is the viscous grain shearing (VGS) model,\citep{Buckingham2007} which takes into account the effect of pore-fluid viscosity on the GS process. Consequently, an additional relaxation time constant is included which helps in better fitting of the experimental data at low frequencies $\left(<10\mbox{ }kHz\right)$, but merges asymptotically with the GS model at high frequencies. Second, the recent VGS($\lambda$) model\citep{Chotiros2010}${}^,$\citep{Buckingham2010} in which the assumption that the effect of pore-fluid viscosity is the same for both the compressional wave and the shear wave is modified, thus yielding two different relaxation time constants. The parameter $\lambda$ represents the wavelength dependence of fluid viscosity in damping of the waves.

\medskip

Buckingham writes the constitutive relation of the modified Maxwell model (see Eqs.~(13)--(16) in Ref.~\citenum{Buckingham2000}) as
\begin{equation}\label{eq03}
\frac{1}{E_{0}}\frac{d\chi}{dt}+\frac{\chi}{\xi\left(t\right)}=\frac{d\varepsilon}{dt}.
\end{equation}
The stress relaxation for the Maxwell element is derived as (see Eqs.~(18) and (19) in Ref.~\citenum{Buckingham2000}):
\begin{equation}\label{eq04}
\chi=\chi_{0}\left(1+\frac{\theta}{\xi_{0}}t\right)^{-E_{0}/\theta},\mbox{ where }\chi_{0}=\left|\varepsilon\right|E_{0}.
\end{equation}
The time-dependent term of Eq.~(\ref{eq04}) along with the introduction of a leading term is identified as the pulse shape function $h\left(t\right)$ and written as:
\begin{equation}\label{eq05}
h\left(t, 0\right)=\frac{\theta}{\xi_{0}}\left(1+\frac{\theta}{\xi_{0}}t\right)^{-E_{0}/\theta}.
\end{equation}
Since the pulse shape function corresponds to a time-varying system, we have added a second temporal argument to it which represents the time of the input relative to the triggering of the time varying viscosity.\citep{Pandey2016} Assuming that this triggering occurs at $t=0$, we will skip the second argument from here onwards in all subsequent steps.

Since the stick-slip processes are randomly distributed across the medium, they are therefore ensemble averaged to obtain the material impulse response functions (MIRFs) $h_{p}\left(t\right)$ and $h_{s}\left(t\right)$ for compressional waves and shear waves respectively. The MIRFs are given as (see Eqs.~(25)--(28) in Ref.~\citenum{Buckingham2000}):
\begin{equation}\label{eq06}
h_{p}\left(t\right)=t_{p}^{-1}\left(1+\frac{t}{t_{p}}\right)^{-\gamma_{p}}\text{, where \ensuremath{t_{p}=\frac{\xi_{0p}}{\theta_{p}}} and \ensuremath{\gamma_{p}=\frac{E_{0p}}{\theta_{p}}}},
\end{equation}
and
\begin{equation}\label{eq07}
h_{s}\left(t\right)=t_{s}^{-1}\left(1+\frac{t}{t_{s}}\right)^{-\gamma_{s}}\text{, where \ensuremath{t_{s}=\frac{\xi_{0s}}{\theta_{s}}} and \ensuremath{\gamma_{s}=\frac{E_{0s}}{\theta_{s}}}}.
\end{equation}
In the above equations, the subscripts ``p'' and ``s'' symbolize the terms associated with pressure or compressional waves and shear waves respectively. Also, the material exponents $n$ and $m$ from Ref.~\citenum{Buckingham2000} are reflected as $\gamma_p$ and $\gamma_s$ respectively in our calculations. This is done to avoid a potential conflict with the symbols reserved for the framework of fractional calculus in Sec.~III. Further, following the argument that the sliding between micro-asperities is indistinguishable for compressional and shear stress relaxations (see Eqs.~(31) and (32) in Ref.~\citenum{Buckingham2000}), we drop the notations of ``p'' and ``s'' from Eqs.~(\ref{eq06}) and (\ref{eq07}) such that
\begin{equation}\label{eq08}
t_{p}=t_{s}=\tau,
\end{equation}
\begin{equation}\label{eq09}
\gamma_{p}=\gamma_{s}=\gamma,
\end{equation}
\begin{equation}\label{eq10}
\mbox{and thus, } h_{p}=h_{s}=h.
\end{equation}
However, it should be noted that the MIRFs for the compressional wave and shear wave are not actually same. This is because the compressional viscoelastic time constant $t_{p}$ and shear viscoelastic time constant $t_{s}$ are not equal, but rather $t_{s}/t_{p}\approx10$. This correction had modified the VGS model to become the VGS($\lambda$) model which gave a better fit to the shear wave dispersion measurements.\citep{Buckingham2007}${}^,$\citep{Buckingham2010} However since the goal of this paper is not to curve-fit the experimental data, but rather show the built-in connection between the GS model and the fractional calculus we will stick to the simplified assumptions represented by Eqs.~(\ref{eq08})--(\ref{eq10}). Moreover, the exact value of the time constants does not alter the form of the wave equations and dispersion relations and therefore, any required scaling could be implemented through a direct substitution in the final expressions.

\medskip

Buckingham then applies the Navier-Stokes equation to study the medium macroscopically resulting in the following two wave equations (see Eqs.~(52) and (53) in Ref.~\citenum{Buckingham2000}):
\begin{equation}\label{eq11}
\nabla^{2}\Psi-\frac{1}{c_{0}^{2}}\frac{\partial^{2}\Psi}{\partial t^{2}}+\frac{\lambda_{p}}{\rho_{0}c_{0}^{2}}\frac{\partial}{\partial t}\nabla^{2}\left[h_{p}\left(t\right)\ast\Psi\right]+\frac{4}{3}\frac{\eta_{s}}{\rho_{0}c_{0}^{2}}\frac{\partial}{\partial t}\nabla^{2}\left[h_{s}\left(t\right)\ast\Psi\right]=0
\end{equation}
and
\begin{equation}\label{eq12}
\frac{\eta_{s}}{\rho_{0}}\nabla^{2}\left[h_{s}\left(t\right)\ast A\right]-\frac{\partial A}{\partial t}=0,
\end{equation}
where $\nabla^{2}$ is the Laplacian operator, $\rho_{0}$ is the bulk density of the material, $c_{0}=\sqrt{E_{0}/\rho_{0}}$ is the lossless phase velocity at zero frequency, and $\lambda_{p}$ and $\eta_{s}$ are stress relaxation coefficients corresponding to compressional waves and shear waves respectively. The terms appearing in the convolution terms of the two Eqs.~(\ref{eq11}) and (\ref{eq12}) are related to the velocity vector $v$ as (see Eq.~(51) in Ref.~\citenum{Buckingham2000}):
\begin{equation}\label{eq13}
v=\nabla\Psi+\nabla\times A.
\end{equation}
The above expression suggests that $\Psi$ and $A$ correspond to the wave displacement field for compressional and shear waves respectively. On taking the Fourier transform of the wave Eqs.~(\ref{eq11}) and (\ref{eq12}), the respective phase velocity $c$ and wave attenuation $\alpha$ is obtained as (see the set of Eqs.~(63) and (64) in Ref.~\citenum{Buckingham2000}):
\begin{equation}\label{eq14}
c_{p}\left(\omega\right)=c_{0}\Re\left[\left\{ 1+\frac{\left(i\omega T\right)^{\gamma}}{\rho_{0}c_{0}^{2}}\left(\varkappa_{p}+\frac{4}{3}\varkappa_{s}\right)\right\} ^{1/2}\right],
\end{equation}
\begin{equation}\label{eq15}
\alpha_{p}\left(\omega\right)=-\frac{\omega}{c_{0}}\Im\left[\left\{ 1+\frac{\left(i\omega T\right)^{\gamma}}{\rho_{0}c_{0}^{2}}\left(\varkappa_{p}+\frac{4}{3}\varkappa_{s}\right)\right\} ^{-1/2}\right]
\end{equation}
for the compressional waves, and
\begin{equation}\label{eq16}
c_{s}\left(\omega\right)=\sqrt{\frac{\varkappa_{s}}{\rho_{0}}}\left(\omega T\right)^{\gamma/2}\sec\left(\gamma\frac{\pi}{4}\right),
\end{equation}
\begin{equation}\label{eq17}
\alpha_{s}\left(\omega\right)=\sqrt{\frac{\rho_{0}}{\varkappa_{s}}}\omega\left(\omega T\right)^{-\gamma/2}\sin\left(\gamma\frac{\pi}{4}\right)
\end{equation}
for the shear waves.
Here, $\Re$ and $\Im$ correspond to the real and imaginary parts of a complex number, and $i=\sqrt{-1}$ is the imaginary number. The terms $\varkappa_{p}$ and and $\varkappa_{s}$ are chosen to represent the compressional and shear rigidity coefficients in order to avoid the conflict with $\gamma$ which is reserved for the material exponent. Further, combining the set of  Eqs.~(62) and (65) in Ref.~\citenum{Buckingham2000}, we can relate the rigidity coefficients with the respective relaxation coefficients as:
\begin{equation}\label{eq18}
\lambda_{p}=\frac{\tau \varkappa_{p}}{\Gamma\left(1-\gamma\right)}\left(\frac{T}{\tau}\right)^{\gamma}
\end{equation}
and
\begin{equation}\label{eq19}
\eta_{s}=\frac{\tau \varkappa_{s}}{\Gamma\left(1-\gamma\right)}\left(\frac{T}{\tau}\right)^{\gamma},
\end{equation}
where $\Gamma\left(\cdot\right)$ is the Euler Gamma function. It must be noted that in the absence of the arbitrary time $T=1 \text{ second}$ introduced by Buckingham in Eqs.~(\ref{eq14})--(\ref{eq19}), the equations become dimensionally inconsistent. The merit of the GS model lies in the fact that besides predicting the dispersion relations for compressional and shear waves as functions of frequency, it also relates to the geo-acoustic parameters of the material, such as grain size, density, viscosity, porosity, and over-burden pressure.\citep{Buckingham2005} However, the material exponent, the time constant, and the stress-relaxation coefficients cannot be deduced from the microscopic properties, but rather heuristically inferred from field measurements of sound speed and attenuation in sediments.\citep{Jackson2007}

\medskip

\section{FRAMEWORK OF FRACTIONAL CALCULUS}

Fractional calculus though as old as the classical Newtonian calculus was rediscovered only lately by Caputo in $1967$ to model the dissipative mechanism in earth materials.\citep{Mainardi2010} Besides offering a mathematical extension to the regular integer order derivatives, fractional derivatives facilitate modeling of materials characterised by spatial and/or temporal memory kernels with arbitrary order exponents. One of the most used forms of the fractional derivative is by Caputo and is defined as the convolution of the power-law memory kernel $\Phi_{m}\left(t\right)$ with the ordinary derivative:\citep{Mainardi2010}
\begin{equation}\label{eq20}
\frac{d^{m}}{dt^{m}}f\left(t\right)\equiv{}_{0}D_{t}^{m}f\left(t\right)\triangleq \Phi_{m}\left(t\right)\ast \left(\frac{d^{n}}{d\tau^{n}}f\left(\tau\right)\right),
\end{equation}
where
\begin{equation}\label{eq21}
\Phi_{m}\left(t\right)=\frac{t^{n-m-1}}{\Gamma\left(n-m\right)}.
\end{equation}
Using Eq.~(\ref{eq21}) in (\ref{eq20}), we have
\begin{equation}\label{eq22}
\frac{d^{m}}{dt^{m}}f\left(t\right)=\frac{1}{\Gamma\left(n-m\right)}\int\limits _{0}^{t}\frac{1}{\left(t-\tau\right)^{m+1-n}}\left(\frac{d^{n}}{d\tau^{n}}f\left(\tau\right)\right)d\tau.\end{equation}
Here $f\left(t\right)$ is a well behaved, causal, continuous function, $n$ is a positive integer, and the real-valued fractional order, $m\in\left(n-1,n\right)$. Further, the Euler Gamma function, $\Gamma\left(\cdot\right)$ is defined for a complex variable $z$ as:
\begin{equation}\label{eq23}
\Gamma\left(z\right)=\int\limits _{0}^{\infty}x^{z-1}e^{-x}dx, \text{ } \Re\left(z\right)>0.
\end{equation}
From Eqs.~(\ref{eq20})--(\ref{eq22}) it can be seen that the power-law memory kernel is built into the fabric of fractional calculus. Thus, fractional calculus offers a compact form of representation for material behavior which is usually described using linear-system theory involving convolutions between integer-order time-derivatives and time-dependent coefficients.

Substituting $n=0$ and $m$ by $-m$ in Eq.~(\ref{eq22}) we obtain the corresponding expression for fractional integral as,
\begin{equation}\label{eq24}
\frac{d^{-m}}{dt^{-m}}f\left(t\right)\equiv{}_{0}I_{t}^{m}f\left(t\right)=\frac{1}{\Gamma\left(m\right)}\int\limits _{0}^{t}\frac{f\left(\tau\right)}{\left(t-\tau\right)^{1-m}}d\tau.\end{equation}

\medskip

Since the Fourier transform of $\Phi_{m}\left(t\right)$ expressed by Eq.~(\ref{eq21}) is a power-law in the frequency domain, it may be even easier to see the extension from the regular integer order derivatives to the fractional order derivatives from:
\begin{equation}\label{eq25}
\mathcal{F}\left[\frac{d^{m}}{dt^{m}}f\left(t\right)\right]=\left(i\omega\right)^{m}F\left(\omega\right),
\end{equation}
where the spatio-temporal Fourier transform is defined as
\begin{equation}\label{eq26}
\mathcal{F}\left[f\left(x,t\right)\right]=F\left(\omega\right)\triangleq\int\limits _{-\infty}^{\infty}\int\limits _{-\infty}^{\infty}f\left(x,t\right)e^{i\left(kx-\omega t\right)}dx\mbox{ }dt,
\end{equation}
where $\omega$ is the temporal angular frequency and $k$ is the corresponding spatial frequency. It should be noted that the choice of positive and negative sign of the kernel gives two definitions of the Fourier transform (see Appendix A of Ref.~\citenum{Holm2014}).

\medskip

\subsection{Kelvin-Voigt fractional derivative wave equation}

Since the Kelvin-Voigt fractional derivative model is central to this paper, we present an illustration of its mechanical equivalent in Fig.~2.

\begin{figure}[H]
\centering
\hspace*{0cm}
\includegraphics[scale=1.5]{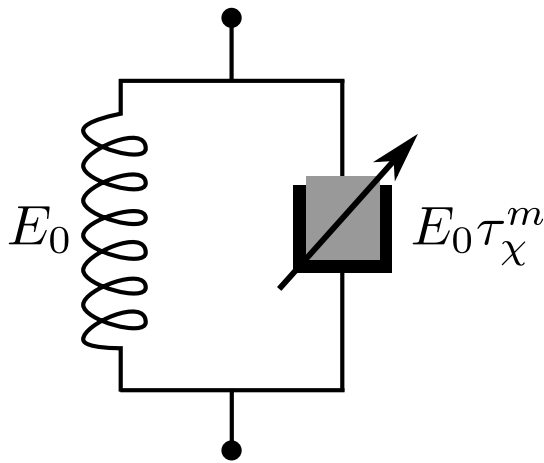}
\caption{(Color online) A mechanical equivalent sketch of the Kelvin-Voigt fractional derivative model comprising a parallel combination of a Hookean spring $E_{0}$, and a fractional viscous dashpot $E_{0}\tau_{\chi}^{m}$.}
\end{figure}

The constitutive stress-strain relation of the model is given as\citep{Mainardi2010}
\begin{equation}\label{eq27}
\chi\left(t\right)=E_{0}\left[\varepsilon\left(t\right)+\tau_{\chi}^{m}\frac{d^{m}\varepsilon\left(t\right)}{dt^{m}}\right].
\end{equation}
The strain when expressed as a time-fractional derivative implies that the material has a long-term memory and remembers its past deformations via a fading memory weighted by a power-law function. The fractional order $m$ is usually in the range from $0$ to $2$ , where $m=1$ corresponds to the standard viscoelastic case. The constitutive Eq.~(\ref{eq27}) when combined with the laws of conservation of mass and momentum which are respectively given as:
\begin{equation}\label{eq28}
\varepsilon\left(t\right)=\frac{\partial}{\partial x}u\left(x,t\right),
\end{equation}
and
\begin{equation}\label{eq29}
\nabla\chi\left(t\right)=\rho_{0}\frac{\partial^{2}}{\partial t^{2}}u\left(x,t\right),
\end{equation}
yields the following wave equation,
\begin{equation}\label{eq30}
\nabla^{2}u-\frac{1}{c_{0}^{2}}\frac{\partial^{2}u}{\partial t^{2}}+\tau_{\chi}^{m}\frac{d^{m}}{dt^{m}}\nabla^{2}u=0,
\end{equation}
where $u\left(x,t\right)$ is the displacement of a plane wave in space $x$ and time $t$. We reserve the discussion of the dispersive characteristics from the Kelvin-Voigt fractional derivative wave equation until next section where we will encounter this equation again.

\medskip

\subsection{Time-fractional diffusion-wave equation}

Ignoring the spring term in Eq.~(\ref{eq27}), we obtain the constitutive relation of the fractional dashpot as,
\begin{equation}\label{eq31}
\chi\left(t\right)=E_{0}\tau_{\chi}^{m}\frac{d^{m}\varepsilon\left(t\right)}{dt^{m}}.
\end{equation}
Now, combining Eq.~(\ref{eq31}) with Eqs.~(\ref{eq28}) and (\ref{eq29}), yields a diffusion-wave equation\citep{Mainardi2010} given as:
\begin{equation}\label{eq32}
\frac{\partial^{2-m}u}{\partial t^{2-m}}=D\frac{\partial^{2}u}{\partial x^{2}}
\end{equation}
where $D=E_{o}\tau_{\chi}^{m}/\rho_{0}>0$, is a constant with the dimension of $L^{2}T^{m-2}$.

On the one hand, in the limit as $m\rightarrow0$, the constitutive Eq.~(\ref{eq31}) corresponds to that of an ideal elastic solid (spring) which is also witnessed from Eq.~(\ref{eq32}) as it approaches a lossless wave equation with $\sqrt{D}$ as the wave velocity. On the other hand, in the limit as $m\rightarrow1$, the spring transforms into the classical Newtonian damper, Eq~(\ref{eq32}) then approaches a standard diffusion equation with $D$ as the diffusion coefficient. For arbitrary values of $m\in\left[0,1\right]$, Eq.~(\ref{eq32}) predicts an interpolation between diffusion and wave like behavior. Inferring physically, the fractional dashpot bridges the gap between a Hookean solid and an ideal Newtonian fluid by intermediating in between them, and therefore it is also termed a spring-pot. Further, if $m\in\left[1,2\right]$, Eq.~(\ref{eq32}) would become a sub-diffusion equation.

\medskip

\section{FROM GS MODEL TO FRACTIONAL WAVE EQUATIONS}

Using  Eqs.~(\ref{eq08})--(\ref{eq10}) we rewrite the expression of MIRF from Eq.~(\ref{eq06}) as,
\begin{equation}\label{eq33}
h\left(t\right)=\tau^{-1}\left(1+\frac{t}{\tau}\right)^{-\gamma}\text{, where \ensuremath{\tau=\frac{\xi_{0}}{\theta}} and \ensuremath{\gamma=\frac{E_{0}}{\theta}}}.
\end{equation}
We stress that Eq.~(\ref{eq33}) shows the long-time inverse power-law behavior similar to the famous Nutting law in rheology.\citep{Mainardi2010} The fractional framework underlying Nutting's law has already been established more than sixty years ago.\citep{Blair1951} The Nutting law is regularly used to curve-fit the non-Debye or non-exponential relaxation processes in complex materials, such as bone, muscles, dielectric materials, polymers and glasses.\citep{Metzler2003} The recognition of the MIRF from the GS model as the Nutting law also suggests that the rheopectic behavior of the pore-fluid is essentially a memory driven non-Markovian process.

\medskip

If the time-varying part of the viscosity dominates, i.e., if $\xi_{0}\ll\theta t$, then the MIRF from Eq.~(\ref{eq33}) can be approximated as, 
\begin{equation}\label{eq34}
h\left(t\right)\sim\tau^{-1}\left(\frac{t}{\tau}\right)^{-\gamma}.
\end{equation}
Here, it should be noted that Buckingham implements the same approximation, but in the frequency domain, see Eqs.~(58) and (59) in Ref.~\citenum{Buckingham2000}. Since, the approximation is better the larger $t$ is relative to $\tau$, the VGS($\lambda$) model actually implies that the approximation is slightly better for compressional waves than for shear waves. Another important observation here is that the approximated MIRF of the time-varying Maxwell model expressed by Eq.~(\ref{eq34}) is equivalent to the relaxation modulus of a fractional dashpot (see Eq.~(3.16b) in Ref.~\citenum{Mainardi2010}). In light of Eqs.~(\ref{eq33}) and (\ref{eq34}), it implies that the elastic $E_{0}$, and the  time-varying viscous properties characterized by $\xi_{0}$ and $\theta$ in the constitutive equation, Eq.~(\ref{eq03}) of the modified Maxwell model jointly manifest themselves to appear as $\tau_{\chi}=\xi_{0}/\theta$ and $m=E_{0}/\theta$ in the constitutive relation of the fractional dashpot given by Eq.~(\ref{eq31}). As noticed, the fractional order $m$ along with the retardation time constant $\tau_{\chi}$ appearing in the fractional framework now gains a physical interpretation. Surprisingly, the observation that the order gives a measure of the interplay between the elastic and viscous properties of the material is in agreement with the empirical observation of  Blair and Coppen\citep{Blair1942} where they referred to the order as the dissipation coefficient, and considered it to be related to the ``firmness'' of the material. Thus, the first direct connection between Buckingham's GS model and fractional calculus is discovered. 

\medskip

\subsection{Compressional wave equation}

Using Eq.~(\ref{eq10}) the last two terms in the compressional wave equation, Eq.~(\ref{eq11}) can be merged together as,
\begin{equation}\label{eq35}
\nabla^{2}\Psi-\frac{1}{c_{0}^{2}}\frac{\partial^{2}\Psi}{\partial t^{2}}+\ensuremath{\left(\frac{\lambda_{p}}{\rho_{0}c_{0}^{2}}+\frac{4}{3}\frac{\eta_{s}}{\rho_{0}c_{0}^{2}}\right)}\frac{\partial}{\partial t}\nabla^{2}\left[h\left(t\right)\ast\Psi\right]=0
\end{equation}
Substituting Eq.~(\ref{eq34}) in Eq.~(\ref{eq35}) we have,
\begin{equation}\label{eq36}
\nabla^{2}\Psi-\frac{1}{c_{0}^{2}}\frac{\partial^{2}\Psi}{\partial t^{2}}+\ensuremath{\left(\frac{\lambda_{p}}{\rho_{0}c_{0}^{2}}+\frac{4}{3}\frac{\eta_{s}}{\rho_{0}c_{0}^{2}}\right)}\tau^{\gamma-1}\frac{\partial}{\partial t}\nabla^{2}\left[ t^{-\gamma}\ast\Psi\right]=0.
\end{equation}
Manipulating the last convolution term as
\begin{equation}\label{eq37}
\left[ t^{-\gamma}\ast\Psi\right]=\Gamma\left(1-\gamma\right)\left[\Psi\ast\frac{t^{-\gamma}}{\Gamma\left(1-\gamma\right)}\right]=\Gamma\left(1-\gamma\right)\left[\frac{d^{1}}{dt^{1}}\left\{ \frac{d^{-1}}{dt^{-1}}\Psi\right\} \ast\frac{t^{-\gamma}}{\Gamma\left(1-\gamma\right)}\right],
\end{equation}
and then comparing Eq.~(\ref{eq37}) in light of Eqs.~(\ref{eq20}) and (\ref{eq21}), we identify $f\left(\tau\right)=d^{-1} \Psi/dt^{-1}$, $n=1$ and $m=\gamma\Rightarrow \gamma\in\left(0, 1\right)$. The constraints in the value of $\gamma$ obtained here in the fractional framework is in agreement with Buckingham\citep{Buckingham2007} where it is shown that for values of  $\gamma\geq1$ the Fourier integrals corresponding to the compressional and shear rigidity coefficients diverge. Alternatively, the same result can also be deduced from the physical constraints imposed on a fractional dashpot. The realizability of a fractional dashpot necessitates that its relaxation function must be positive and completely monotonically decreasing which can only be satisfied for $\gamma<1$.\citep{Mainardi2010} This finding also corresponds to the thermodynamic constraint imposed on a fractional dashpot by the Clausius--Duhem inequality.\citep{Lion1997}

\medskip

The term with a negative fractional order time derivative in Eq.~(\ref{eq37}) is equivalent to the fractional integration given by Eq.~(\ref{eq24}). Using Eqs.~(\ref{eq20}), (\ref{eq21}) and (\ref{eq37}), the fractional order derivative equivalent of the convolution term in Eq.~(\ref{eq36}) can be written as,
\begin{equation}\label{eq38}
\left[ t^{-\gamma}\ast\Psi\right]=\Gamma\left(1-\gamma\right)\frac{d^{\gamma-1}}{dt^{\gamma-1}}\Psi.
\end{equation}
Substituting Eq.~(\ref{eq38}) back in Eq.~(\ref{eq36}) and rearranging the terms we get,
\begin{equation}\label{eq39}
\nabla^{2}\Psi-\frac{1}{c_{0}^{2}}\frac{\partial^{2}\Psi}{\partial t^{2}}+\Gamma\left(1-\gamma\right)\ensuremath{\left(\frac{\lambda_{p}}{\rho_{0}c_{0}^{2}}+\frac{4}{3}\frac{\eta_{s}}{\rho_{0}c_{0}^{2}}\right)\tau^{\gamma-1}}\frac{d^{\gamma}}{dt^{\gamma}}\nabla^{2}\Psi=0.
\end{equation}
We find that Eq.~(\ref{eq39}) is equivalent to the Kelvin-Voigt fractional derivative wave Eq.~(\ref{eq30}), where
\begin{equation}\label{eq40}
\tau_{\chi}^{\gamma}=\Gamma\left(1-\gamma\right)\ensuremath{\left(\frac{\lambda_{p}}{\rho_{0}c_{0}^{2}}+\frac{4}{3}\frac{\eta_{s}}{\rho_{0}c_{0}^{2}}\right)\tau^{\gamma-1}}.
\end{equation}
The dimensional consistency of Eq.~(\ref{eq40}) suggests the validity of  Eq.~(\ref{eq39}), and hence the mapping of the compressional wave equation, Eq.~(\ref{eq11}) of the GS model into the fractional framework. Also, from Eq.~(\ref{eq40}) we obtain the relationship between the characteristic relaxation time constant of a material with its geo-acoustic parameters though there are limitations in the determination of relaxation coefficients as mentioned earlier. It is interesting to observe that though the MIRF from the GS model is found similar to the relaxation modulus of the fractional dashpot alone, the resulting wave equation from the GS model turns out to be that of a Kelvin-Voigt fractional derivative model. The reason can be traced back to Eqs.~(37) and (44) in Ref.~\citenum{Buckingham2000}, where the acoustic pressure corresponding to a suspension without any grain-to-grain stress relaxation, is included in the expression of the stress tensor.  Such a loss-less medium is represented by a spring. Since stresses add up in parallel branches of viscoelastic models, it implies that the GS mechanism has effectively been modeled using a parallel combination of a spring and a time-varying Maxwell model. Further, as already mentioned, the relaxation modulus of the time-varying Maxwell model is equivalent to that of a fractional dashpot. Thus, the Kelvin-Voigt fractional derivative wave equation obtained here is quite natural. 

\medskip

On the one hand, in the limiting case of maximum rheopectic behavior, i.e. if $\theta\rightarrow\infty\Rightarrow\gamma\rightarrow0$, Eq.~(\ref{eq39}) then approaches the familiar lossless wave equation. Physically, it implies that the intergranular sliding has stopped which is possible if grains are locked against each other. Such a situation is plausible, at least locally among the participating grains, if the pore-fluid is completely squeezed out as a result of the intergranular sliding. In an ideal condition, the initially assumed unconsolidated granular material would then effectively transform into a compact solid and therefore any possible energy dissipation in the form of a viscous drag force would be ruled out. On the other hand, as the rheopectic behavior decreases, i.e. $\gamma$ increases, grain-shearing is facilitated and attenuation rises. Besides, in such cases varying degrees of flow of the pore-fluid in between the grains cannot be neglected. In the limit as $\gamma\rightarrow1$, the wave Eq.~(\ref{eq39}) approaches the classical viscous wave equation. 

The dispersive behavior from the Kelvin-Voigt fractional derivative wave Eq.~(\ref{eq30}) has already been studied in detail.\citep{Holm2010}${}^-$\citep{Holm2013} However, for the sake of completeness, we repeat the necessary mathematical framework which will be utilized in the next subsection when we analyze the shear wave equation.

For modeling of dispersive properties in a material, we assume the wave propagation vector $k$ to be complex such that
\begin{equation}\label{eq41}
k=\beta-i\alpha, \text{ } \beta\geq0 \text{ and } \alpha\geq0,
\end{equation}
where $\beta$ is the wave velocity vector and $\alpha$ is the wave attenuation vector. Both $\beta$ and $\alpha$ are functions of $\omega$, and also related to each other by the Kramers-Kronig relations due to causality.\citep{Carcione2001} Since, the Kramers-Kronig relations are fundamentally the Hilbert transform pair, it also couples attenuation of the wave with its velocity. 

We further assume a unit amplitude, one-dimensional, plane wave whose displacement is given as $\Psi\left(x,t\right)=e^{i\left(\omega t-kx\right)}$, is incident on the medium. Then using Eq.~(\ref{eq41}), we obtain the expression for a propagating wave field in a lossy medium as,
\begin{equation}\label{eq42}
\Psi\left(x,t\right)=e^{-\alpha x}e^{i\left(\omega t-\beta x\right)}.
\end{equation}
On taking the Fourier transform of the wave Eq.~(\ref{eq39}), and using Eq.~(\ref{eq25}), we obtain the dispersion relation as
\begin{equation}\label{eq43}
k=\frac{\omega}{c_{0}}\sqrt{\frac{1}{1+\left(i\omega\tau_{\chi}\right)^{\gamma}}}
\end{equation}
where $\tau^{\gamma}_{\chi}$ is given by Eq.~(\ref{eq40}). As can be seen in the limit as $\gamma\rightarrow0$, Eq.~(\ref{eq43}) approaches the lossless dispersion relation. On the contrary, in the limit as $\gamma\rightarrow1$, Eq.~(\ref{eq43}) approaches the classical damped viscous-wave dispersion relation.

\medskip

The simple closed-form dispersion Eq.~(\ref{eq43}) apparently looks different from the respective dispersion relations from the GS model expressed by Eqs.~(\ref{eq14}) and (\ref{eq15}), but it is not. Substituting the values of stress-relaxation coefficients from Eqs.~(\ref{eq18}) and (\ref{eq19}) in Eq.~(\ref{eq40}) gives,
\begin{equation}\label{eq44}
\tau_{\chi}=T\left[\frac{1}{\rho_{0}c_{0}^{2}}\left(\varkappa_{p}+\frac{4}{3}\varkappa_{s}\right)\right]^{\frac{1}{\gamma}},
\end{equation}
which, when substituted in Eq.~(\ref{eq43}) yields,
\begin{equation}\label{eq45}
k=\frac{\omega}{c_{0}}\left[1+\frac{\left(i\omega T\right)^{\gamma}}{\rho_{0}c_{0}^{2}}\left(\varkappa_{p}+\frac{4}{3}\varkappa_{s}\right)\right]^{-\frac{1}{2}}.
\end{equation}
The real part of the wave propagation vector, $\beta_{p}$ from Eq.~(\ref{eq45}), when substituted in the expression of phase velocity, $c_{p}\left(\omega\right)=\omega/\beta_{p}$ results into Eq.~(\ref{eq14}). Similarly, the imaginary part of the wave propagation vector, $\alpha_{p}$ gives the wave attenuation expressed by Eq.~(\ref{eq15}). As evident, the dispersion relations obtained for the compressional wave in the fractional framework are essentially the same as those obtained in the GS model. The equivalence of the dispersion relations in the frequency domain dictates the corresponding equivalence of the respective wave equations in the time domain. Thus, Buckingham's compressional wave equation, Eq.~(\ref{eq11}) from the GS model, and the Kelvin-Voigt fractional-derivative wave equation, Eq.~(\ref{eq39}), are two different yet mathematically equivalent expressions. In the next subsection we will demonstrate the equivalence for shear waves as well. Further, the observations made from curve-fitting of the experimental data\citep{Buckingham2000}${}^,$\citep{Buckingham2007} with the dispersion relations from the GS model are equally applicable for the dispersion relations derived in the fractional work. Therefore, we choose not to repeat the same curve-fitting exercise already done before, but rather show the benefits of the fractional framework.

\medskip

It should be noted that contrary to the assumptions followed in the GS model, and hence in the derivation of Eq.~(\ref{eq45}), there is viscous dissipation in the pore-fluid. The VGS model which takes into account the viscous dissipation, is characterized by an additional relaxation time constant. This also reflects in the form of an additional dissipation term in the respective expressions of phase velocity and wave attenuation for both compressional waves and shear waves, see Eqs.~(34)--(39) in Ref.~\citenum{Buckingham2007}. Further, the values of the material exponent and rigidity coefficients in the GS model differ slightly from those in the VGS model, see Table I in Ref.~\citenum{Buckingham2007}. Furthermore, the recent VGS($\lambda$) model takes into account that the viscoelastic time constants for compressional and shear waves are not equal. Consequently, the VGS($\lambda$) model gives a better fit to the shear wave dispersion measurements.\cite{Buckingham2010}

\medskip

The asymptotic trends of the phase velocity and wave attenuation are obtained following simple steps which then facilitate a quick and qualitative interpretation of the wave dispersion in a given medium.\citep{Holm2010} Summarizing them here:
\begin{equation}\label{eq46}
c_{p}\left(\omega\right)\begin{cases}
=c_{0}, & \left(\omega\tau_{\chi}\right)^{\gamma}\ll1\\
\propto\omega^{\frac{\gamma}{2}}, & 1\ll\left(\omega\tau_{\chi}\right)^{\gamma}
\end{cases}
\end{equation}
and 
\begin{equation}\label{eq47}
\alpha_{p}\left(\omega\right)\begin{cases}
\propto\omega^{1+\gamma}, & \left(\omega\tau_{\chi}\right)^{\gamma}\ll1\\
\propto\omega^{1-\frac{\gamma}{2}}, & 1\ll\left(\omega\tau_{\chi}\right)^{\gamma}.
\end{cases}
\end{equation}
The conditions $\left(\omega\tau_{\chi}\right)^{\gamma}\ll1$ and $1\ll\left(\omega\tau_{\chi}\right)^{\gamma}$ mentioned in Eqs.~(\ref{eq46}) and (\ref{eq47}) characterize the low frequency and high frequency regimes respectively.

\medskip

We set $c_{0}=1$ and $\tau_{\chi}=1$ for plotting purpose. The phase velocity dispersion curve and attenuation curve for different values of $\gamma$ are shown in Fig.~3 and Fig.~4 respectively, where the markers; square for $\gamma=0.3$, circle for $\gamma=0.5$ and triangle for $\gamma=0.9$ represent the upper cut-off of the dispersion curve corresponding to a penetration depth of one wavelength. The marker for $\gamma=0.1$ lies outside the given frequency range. For visualization convenience, each attenuation curve has been normalized to its value at $\omega  \tau_{\chi}=1$. Further, the gradual increase of the phase velocity, and the near-linear power-law attenuation of the compressional wave with respect to frequency can be observed from the dispersion curve corresponding to $\gamma=0.1$ in Fig.~3 and Fig.~4 respectively.  This observation regarding small values of $\gamma$ are the most interesting ones since such values are used by Buckingham for the curve-fitting of experimental data, see Table I and Fig.~7 in Ref.~\citenum{Buckingham2007}. 

\begin{figure}[H]
\centering
\hspace*{0cm}
\includegraphics[scale=0.38]{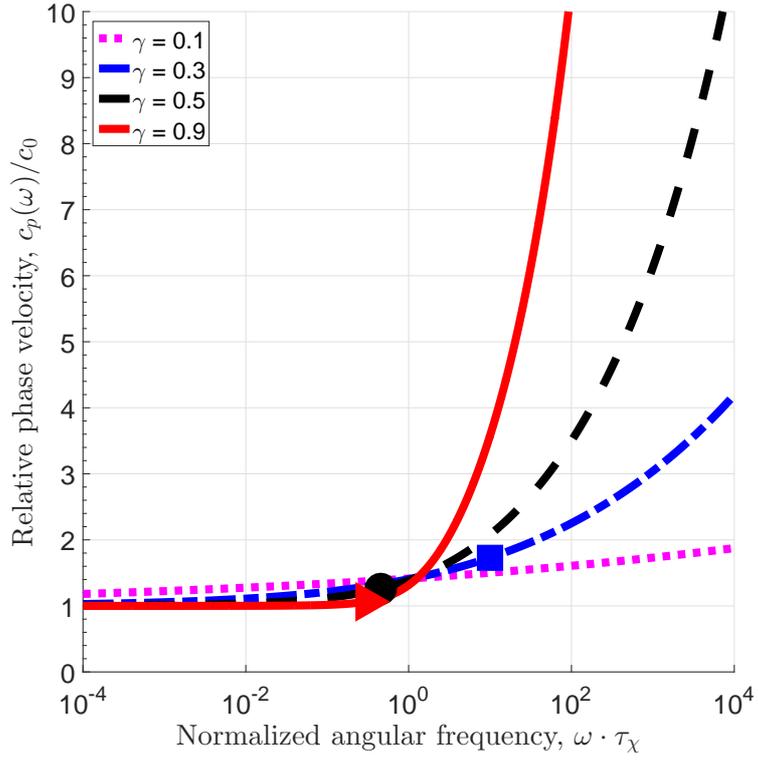}
\caption{(Color online) Frequency-dependent phase velocity dispersion
for the compressional waves from the GS model. The fractional derivative order $\gamma$ has values 0.1 (dotted line), 0.3 (dash-dot line), 0.5 (dashed line) and 0.9 (solid line).}
\end{figure}

\begin{figure}[H]
\centering
\hspace*{0cm}
\includegraphics[scale=0.38]{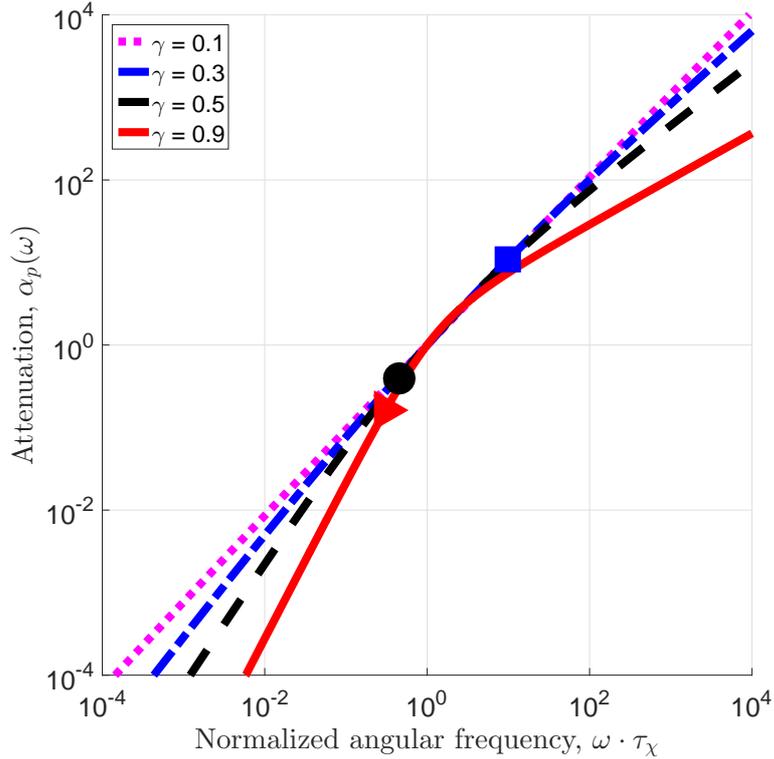}
\caption{(Color online) Frequency-dependent wave attenuation for the compressional waves from the GS model. The fractional derivative order $\gamma$ has values 0.1 (dotted line), 0.3 (dash-dot line), 0.5 (dashed line) and 0.9 (solid line).}
\end{figure}

As shown in Fig.~3, and also seen from the asymptotic behavior expressed by Eq.~(\ref{eq46}), for larger values of $\gamma$, as $\omega\rightarrow\infty$, the phase velocity $c_{p}\rightarrow\infty$. This apparently looks like a violation of causality however it is not, since as $\omega\rightarrow\infty$, attenuation $\alpha_{p}\rightarrow\infty$, see Eq.~(\ref{eq47}) and Fig.~4. Moreover, it can be seen from Eqs.~(\ref{eq46}) and (\ref{eq47}) that the rate of increase of attenuation is comparatively higher than that of the phase velocity for the same frequency values. This aspect of the dispersive behavior can be better explained by employing the notion of penetration depth or skin depth which is often used in electromagnetic studies. The penetration depth $\delta$ of the propagating wave is defined as the distance traversed by the wave before its amplitude falls by a factory of $1/e\approx36.7\mbox{ }\%$, or intensity becomes $1/e^{2}\approx13.5\mbox{ }\%$ of its maximum value.\citep{Ziomek1994} Imposing this condition on Eq.~(\ref{eq42}) gives,
\begin{equation}\label{eq48}
\delta\left(\omega\right)=\frac{1}{\alpha\left(\omega\right)}.
\end{equation}
The penetration depth $\delta_{p}$ for compressional waves can be better quantified in number of wavecycles as,
\begin{equation}\label{eq49}
\delta_{p}\left(\omega\right)=\frac{\omega}{2\pi\alpha_{p}\left(\omega\right)c_{p}\left(\omega\right)}\mbox{ (in number of wavelengths).} 
\end{equation}
Using Eqs.~(\ref{eq46}), (\ref{eq47}) and (\ref{eq49}) we obtain the asymptotic behavior of the penetration depth for compressional waves in low frequency and high frequency regimes as:
\begin{equation}\label{eq50}
\delta_{p}\left(\omega\right)\begin{cases}
\propto\omega^{-\gamma}, & \left(\omega\tau_{\chi}\right)^{\gamma}\ll1\\
=\text{constant}, & 1\ll\left(\omega\tau_{\chi}\right)^{\gamma}.
\end{cases}
\end{equation}
As shown in Fig.~5, the value of the penetration depth is greater for low frequencies. With increase in frequency, the penetration depth obeying power-law given by Eq.~(\ref{eq50}) falls and approaches a constant value for high frequencies. In the case of viscous media, i.e. for larger values of $\gamma$, the penetration depth falls even more rapidly and is reduced to less than a single wavelength in the high frequency regime. Physically, it implies that the wave enters the evanescent mode where its oscillatory motion ceases to exist and the wave finally decays. This is consistent with Buckingham's observation regarding Stokes' equation $\left(\gamma=1\right)$ that an infinitely fast wave suffers an infinite attenuation, and thus its propagation distance in the medium is zero.\citep{Buckingham2005b} Interestingly this behavior is not limited to $\gamma=1$, but rather as seen from Fig.~5, propagation modes corresponding to all other permitted values of $\gamma$ also undergo infinite attenuation in the high frequency regime and become evanescent.

\begin{figure}[H]
\centering
\hspace*{0cm}
\includegraphics[scale=0.38]{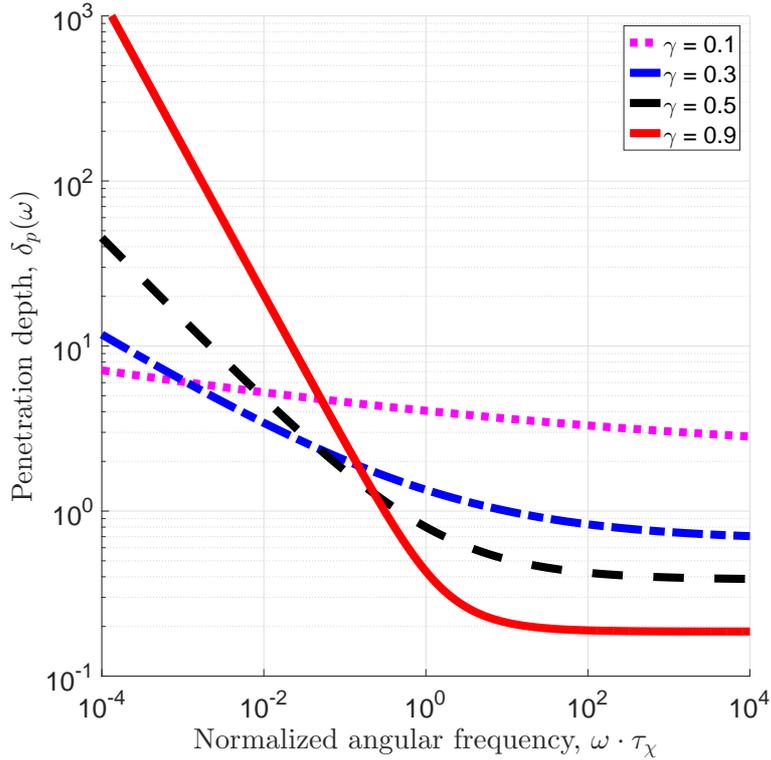}
\caption{(Color online) Frequency-dependent penetration depth (in
number of wavelengths) for the compressional waves from the
GS model. The fractional derivative order $\gamma$ has values 0.1 (dotted line), 0.3 (dash-dot line), 0.5 (dashed line) and 0.9 (solid line).}
\end{figure}

\medskip

\subsection{Shear wave equation}

Replacing the wave displacement field $\Psi$ by $A$ in Eq.~(\ref{eq38}) and then substituting it along with Eq.~(\ref{eq34}) in Eq.~(\ref{eq12}), we get
\begin{equation}\label{eq51}
\Gamma\left(1-\gamma\right)\frac{\eta_{s}}{\rho_{0}}\tau{}^{\gamma-1}\nabla^{2}\left[\frac{d^{\gamma-1}}{dt^{\gamma-1}}A\right]-\frac{\partial A}{\partial t}=0.
\end{equation}
On further simplification Eq.~(\ref{eq51}) becomes,
\begin{equation}\label{eq52}
\frac{\partial^{2-\gamma}A}{\partial t^{2-\gamma}}=\Gamma\left(1-\gamma\right)\frac{\eta_{s}}{\rho_{0}}\tau^{\gamma-1}\nabla^{2}A.
\end{equation}
Comparing Eq.~(\ref{eq52}) with Eq.~(\ref{eq32}), we find that the shear wave equation from the GS model is equivalent to a time-fractional diffusion-wave equation, where the constant $D$ takes the form, 
\begin{equation}\label{eq53}
D=\frac{\eta_{s}}{\rho_{0}}\tau^{\gamma-1}\Gamma\left(1-\gamma\right).
\end{equation}
An even simpler expression is obtained by substituting Eq.~(\ref{eq19}) in Eq.~(\ref{eq53}) as,
\begin{equation}\label{eq54}
D=\frac{\varkappa_{s}}{\rho_{0}} T^{\gamma}.
\end{equation}
Then Eq.~(\ref{eq52}) attains its final form:
\begin{equation}\label{eq55}
\frac{\partial^{2-\gamma}A}{\partial t^{2-\gamma}}=D\nabla^{2}A.
\end{equation}
Since $0<\gamma<1$, the fractional order in Eq.~(\ref{eq55}) is, $1<\left(2-\gamma\right)<2$. The time domain solution of the equation suggests that as $\gamma$ decreases from $1$ to $0$, the phenomenon of diffusion transforms into a lossless wave propagation. Following the same approach as in the case of the compressional wave equation, the Fourier transform of Eq.~(\ref{eq55}) is taken to obtain the following dispersion relation,
\begin{equation}\label{eq56}
k=\frac{\omega^{1-\frac{\gamma}{2}}}{\sqrt{D}} i^{-\frac{\gamma}{2}}=\frac{\omega^{1-\frac{\gamma}{2}}}{\sqrt{D}}  \left\{ \cos\left(\gamma\frac{\pi}{4}\right)-i\mbox{ }\sin\left(\gamma\frac{\pi}{4}\right)\right\}.
\end{equation}
Extracting the real and imaginary parts of Eq.~(\ref{eq56}), we obtain the required expressions for phase velocity and attenuation of shear waves as:
\begin{equation}\label{eq57}
c_{s}\left(\omega\right)=\frac{\omega}{\beta_{s}\left(\omega\right)}=\sqrt{D}\omega^{\frac{\gamma}{2}}\sec\left(\gamma\frac{\pi}{4}\right)
\end{equation}
and
\begin{equation}\label{eq58}
\alpha_{s}\left(\omega\right)=\frac{\omega^{1-\frac{\gamma}{2}}}{\sqrt{D}} \sin\left(\gamma\frac{\pi}{4}\right).
\end{equation}
On comparing Eqs.~(\ref{eq57}) and (\ref{eq58}) with Eqs.~(\ref{eq46}) and (\ref{eq47}) respectively, we observe that in contrast to the compressional wave dispersion relations which are characterized by two different power-laws, the shear wave dispersion relations are characterized by a single power-law in the entire frequency regime. Also, the power-law behavior of phase velocity and attenuation of shear waves is similar to the respective asymptotic behavior of phase velocity and attenuation of compressional waves in the high frequency regime. The dispersion relations for the shear waves in the GS model expressed by Eqs.~(\ref{eq16}) and (\ref{eq17}) can be recovered by a direct substitution of Eq.~(\ref{eq54}) in Eqs.~(\ref{eq57})  and (\ref{eq58}). Thus, the equivalence of the shear wave equation, Eq.~(\ref{eq12}) from the GS model, and the fractional diffusion-wave equation, Eq.~(\ref{eq55}), is established. As expected, the shear wave dispersion equations, Eqs.~(\ref{eq57}) and (\ref{eq58}) are independent of the term $c_{0}$ which is attributed to the lossless spring in the Kelvin-Voigt fractional derivative model.

\medskip

Further, as observed from Eqs.~(\ref{eq57}) and (\ref{eq58}), the phase velocity vector, $\beta_{s}$, and the attenuation vector, $\alpha_{s}$, follow the same power-law dependence on frequency. The equality of the competing race between the two vectors can also be witnessed in the frequency independent expression of the penetration depth for shear waves, which is obtained as,
\begin{equation}\label{eq59}
\delta_{s}=\frac{\omega}{2\pi\alpha_{s}\left(\omega\right)c_{s}\left(\omega\right)}=\frac{1}{2\pi}\cot\left(\gamma\frac{\pi}{4}\right) \mbox{(in number of wavelengths)}.
\end{equation}
On the one hand, from Eq.~(\ref{eq59}), in the limit as $\gamma\rightarrow0$, $\delta_{p}\rightarrow\infty$, physically implies the classical lossless wave propagation. On the other hand, if $\gamma\rightarrow1$, then from Eq.~(\ref{eq57}) as $\omega\rightarrow\infty$, $c_{s}\rightarrow\infty$, however the penetration depth given by Eq.~(\ref{eq59}) is restricted to less than a single wavelength. This corresponds to the characteristic evanescent wave solution expected from the standard diffusion equation. Thus, similar to the case for compressional waves in the GS model, causality is ensured for shear waves as well.

\medskip

We set $D=1$ for the plotting purpose. As seen from Eq.~(\ref{eq59}), shear waves become evanescent, i.e., $\delta_{s}\leq1$ for $\gamma \geq 0.2$. Therefore, very small values of $\gamma$ are chosen for the plotting of phase velocity dispersion curves for shear waves in Fig.~6.

\begin{figure}[H]
\centering
\hspace*{0cm}
\includegraphics[scale=0.38]{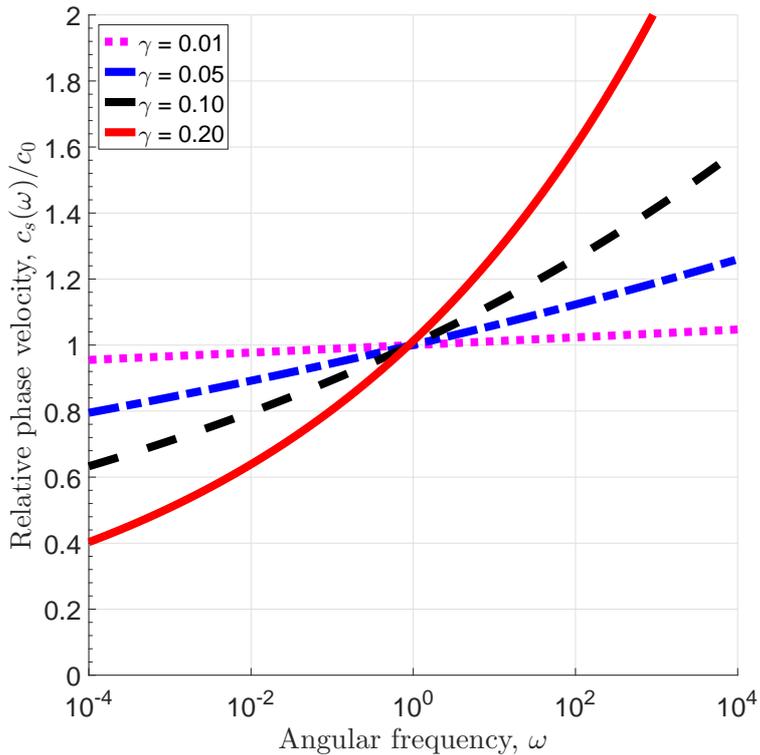}
\caption{(Color online) Frequency-dependent phase velocity dispersion
for the shear waves from the GS model. The fractional derivative order $\gamma$ has values 0.01 (dotted line), 0.05 (dash-dot line), 0.10 (dashed line) and 0.20 (solid line).}
\end{figure}

Further, the wave attenuation expressed by Eq.~(\ref{eq58}) does not change significantly for the propagating shear waves, i.e., for $\gamma < 0.2$. Since for such small values of $\gamma$, the slope of the attenuation--frequency curve hardly deviates from $1$, the attenuation curves for shear waves are not plotted as then it would be trivial. 

\medskip

\section{DISCUSSION AND CONCLUSION}

On the one hand, the two disparate fields of sediment acoustics and non-Newtonian rheology have traditionally developed nearly independent of each other. It is partially because the intended goals are different in the two fields. Also, the experimental instruments employed are not usually common to the two fields. On the other hand, fractional calculus has developed as a standalone mathematical framework, lacking a proper connection to real physics. There are mainly two goals achieved in this paper. First, this may be the first result which directly connects the fractional framework to a process deeply rooted in physics. We have shown that the equations derived from the fundamental physical process of grain-shearing could also lead to fractional order wave equations. The connection is further validated from the dispersion results in the fractional framework which are shown to be equivalent to their respective counterparts in the GS model.  A bonus of this bridging is the physical significance of the fractional order which is extracted from the GS model. An estimation of the order from the physical parameters of the material would now greatly reduce the ambiguities in the curve-fitting of the experimental data. Besides, this will also help in identifying the underlying physical processes in materials which exhibit power-law jumps from one form to another, such as, from a damped diffusive process to a propagating wave. With increase in complexity of the material, there is a demand for a more flexible but yet a robust mathematical framework for their modeling. Fractional calculus is one of the candidates which could meet this demand. Moreover, without adopting a fractional calculus approach, such complex material behavior could be difficult to model, both analytically as well as numerically.\citep{Holm2015} 

\medskip

Second, the term ``strain-hardening'' which is central to the GS model has been recognized as the property of rheopecty. Such hardening behavior has already been shown to result in the Lomnitz logarithmic creep law,\citep{Pandey2016} the seismological applications of which can be witnessed in rock physics,\citep{Lomnitz1956} marine sediments\citep{Stoll1970} and earthquake modeling.\citep{Charpentier2015} By establishing the equivalence of wave equations and dispersion relations of the GS model to the respective equations obtained in the fractional framework we have also demonstrated the possibility of modeling time-dependent non-Newtonian fluid properties using fractional derivatives. Such fluid properties have remained an open problem from modeling perspectives so far. Besides, the similarity between a fractional dashpot and a time-varying Maxwell model would now encourage the use of fractional viscoelastic models in both the fields of sediment acoustics as well as non-Newtonian rheology.\citep{Pandey2016} The bridging of the two fields is expected to be symbiotic.
\medskip

Another possible advantage from this bridging is that the dispersion relations in the fractional framework are solely determined by the characteristic retardation time constant $\tau_{\chi}=\xi_{0}/\theta$, and the fractional order $\gamma=E_{0}/\theta$, which although challenging could be measured using the experimental techniques of non-Newtonian rheology.\citep{Dewar2006} The time-dependent fluid behavior can be measured by a more direct steady shear test as well as by the hysteresis loop test. This could then possibly free the GS model from the heuristic estimation of curve-fitting parameters. Further work should include an investigation of the VGS model in the framework of fractional calculus and examine if it is as consistent as the GS model in the fractional framework.

\medskip

 \setlength{\parindent}{0.7cm}

 \smallskip
\vspace{50pt}

\textcolor{blue}{\textbf{Please cite to this paper as published in:}}

\vspace{20pt}

V. Pandey and S. Holm, ``Connecting the grain-shearing mechanism of wave propagation in marine sediments to fractional order wave equations,'' J. Acoust. Soc. Am. \textbf{140} (6), 4225-4236 (2016).
\vspace{30pt}

\textbf{ \textit{DOI: \href{http://dx.doi.org/10.1121/1.4971289}{http://dx.doi.org/10.1121/1.4971289}}}

 \end{document}